\documentclass[twocolumn,preprintnumbers,amsmath,amssymb]{revtex4}
\usepackage{graphicx}
\usepackage{dcolumn}
\usepackage{bm}

\begin{document}

\title{Control of Four-Level Quantum Coherence via Discrete Spectral Shaping of an Optical Frequency Comb}

\author{Matthew C. Stowe, Avi Pe'er, and Jun Ye}
 \affiliation{JILA, National Institute of Standards and Technology and University of Colorado\\Department of Physics, University of Colorado, Boulder, CO 80309-0440}

\date{\today}

\begin{abstract}
We present an experiment demonstrating high-resolution coherent
control of a four-level atomic system in a closed (diamond) type
configuration. A femtosecond frequency comb is used to establish
phase coherence between a pair of two-photon transitions in cold
$^{87}$Rb atoms. By controlling the spectral phase of the frequency
comb we demonstrate the optical phase sensitive response of the
diamond system. The high-resolution state selectivity of the comb is
used to demonstrate the importance of the signs of dipole moment
matrix elements in this type of closed-loop excitation. Finally, the
pulse shape is optimized resulting in a 256\% increase in the
two-photon transition rate by forcing constructive interference
between the mode pairs detuned from an intermediate resonance.

\end{abstract}

\maketitle The field of coherent control of light-matter
interactions for the purpose of driving a quantum system to a
desired state has been drawing increasing interest. Research in the
field of coherent control generally uses either an ultrashort pulse
to create atomic coherences over a very large bandwidth, or in the
other extreme, continuous wave (cw) lasers to drive transitions with
high-resolution. Some of the pioneering work on coherent control
used pulse shaping of a single broad-bandwidth pulse to enhance or
diminish two-photon absorption \cite{Noordam1,Dudovich1}, improve
the resolution of coherent anti-Stokes Raman scattering
\cite{Oron1}, and control molecular wavepacket motion
\cite{Salzmann1}. Simultaneously, a tremendous amount of research
has been done using narrow-bandwidth cw-lasers to control
three-level atomic systems, for example to study electromagnetic
induced transparency \cite{Fleischhauer1}. Particularly relevant to
this Letter are the many theoretical studies of four-level systems,
for example in a double-Lambda \cite{Korsunsky1} or diamond type
configuration \cite{Morigi1,Lukin1}. The advent of optical frequency
combs has made a strong impact on the field of high-precision
spectroscopy \cite{Stowe1}, and has recently been used for coherent
control of a three-level transition \cite{Stowe2}. In this work we
combine the broad-bandwidth and high-resolution of the comb,
augmented by discrete spectral phase shaping, to control the phase
sensitive response of a four-level atomic system in a diamond
configuration.

We use a femtosecond optical frequency comb to excite a pair of
resonant two-photon transitions that form a closed diamond
configuration in cold $^{87}$Rb. The broad-bandwidth of the
femtosecond pulses allows for a two-photon transition to be excited
via different intermediate states with a separation of 7
THz. Simultaneously the narrow linewidth of each comb mode allows
the excitation of only specific atomic levels, a necessary condition
to isolate the four-level diamond configuration from the full set of
atomic transitions. The phase sensitive excitation of this closed
four-level system is demonstrated by discrete spectral shaping of
the femtosecond comb. The 7 THz separation between intermediate
states allows use of a pulse shaper to change the phase
of the comb mode that is resonant with one intermediate state and
not the other resonant modes, this shifts the relative phase between
the two paths constituting the diamond. By analogy with an optical
interferometer, the two-photon excitation is shown to exhibit
a sinusoidal variation versus the applied phase, due to the resonant comb modes alone. In the second experiment the
off-resonant comb modes are of primary importance, in particular we show that the two-photon excitation is enhanced by
changing the phase of the two-photon resonant mode pairs that are symmetrically detuned from an intermediate state.

The experiments presented in this Letter can be modeled as the
interaction of a phase coherent train of shaped femtosecond pulses
(an optical comb) with a four-level atomic system in a diamond
configuration. A diamond configuration has one ground state, two
intermediate states, and a single excited state such that there are
a total of four possible resonant dipole allowed transitions. There
are hundreds of thousands of frequency modes in the comb spectrum,
four of which are tuned to be resonant with the transitions
shown in Fig. 1(a). Due to the equidistant spacing of comb
frequencies, any mode in the spectrum has another mode that forms a
two-photon resonant pair. Therefore there are hundreds of thousands
of mode pairs that are two-photon resonant but have varying
detunings from the intermediate states, these will be referred to as
off-resonant mode pairs. We present our experiments in two parts,
the first focuses on the diamond configuration excited by resonant
modes only, the second part utilizes another pulse shape to enhance
the signal from the off-resonant mode pairs.

Two-photon absorption via a pair of modes and a single intermediate
state gives rise to an excited state amplitude that within
second-order perturbation is given by \cite{Yoon1},
\begin{eqnarray}
\nonumber c_{gf}\propto \frac{E_nE_m\mu_{gi}\mu_{if}}{i(\omega_{gf}-(m+n)2\pi f_r-4\pi f_o)+\pi \gamma_{f}}\times \\
\nonumber \left[\frac{1}{i(\omega_{gi}-2\pi(nf_r+f_o))+\pi \gamma_{i}}\right.+ \\
\left.\frac{1}{i(\omega_{gi}-2\pi(mf_r+f_o))+\pi \gamma_{i}}\right],
\end{eqnarray}\label{TwoPhotonFormula}
where $E_{n,m}$ are the electric fields of the $n^{th}$ and $m^{th}$
modes of the comb, $\gamma_{i(f)}$ is the intermediate (final) state
decay rate, $\omega_{gi(gf)}$ is ground to intermediate (final)
state transition frequency, $f_r$ is the repetition frequency of the
comb, $f_o$ is the comb offset frequency, and $\mu_{gi(if)}$ are the
dipole moments from the ground to intermediate (intermediate to
final) states. The total excited state amplitude is then given by
the sum of all the possible two-photon resonant transition pathways
resulting from all comb mode pairs in the laser spectrum connecting
through various intermediate states. The key physics for the results
presented here is that the phase of the excited state amplitude is a
function of the detuning from the intermediate state, the signs of
dipole moment matrix elements, and the phase of the two electric
fields. In particular, the phase of the excited state amplitude is
+90$^o$(-90$^o$) for detunings above(below) the intermediate state, and is 0$^o$ for zero intermediate state detuning. Due
to the phase difference of 180$^o$ between two excited state
amplitudes symmetrically detuned about an intermediate resonance,
the contribution to the total amplitude from the off-resonant mode
pairs cancels to zero for a train of transform-limited pulses.

\begin{figure}[h]
\resizebox{8.5cm}{!}{
\includegraphics[angle=0]{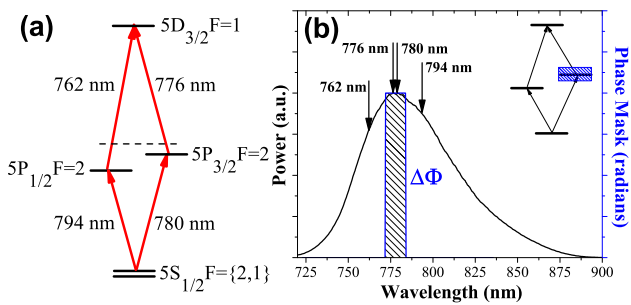}}
\caption{\label{Fig1}$\bf(a)$ Energy level diagram of the diamond
configuration in $^{87}$Rb, the arrows indicate the resonant comb
modes. $\bf(b)$ Pulse spectrum and phase mask used for the first
experiment. The hatched region in the spectrum and the inset energy
level diagram indicates the portion of the spectrum to which the
phase mask is applied. The four arrows indicate resonant
wavelengths in relation to the phase mask spectral window.}
\end{figure}

The experiments are conducted on an ensemble of cold $^{87}$Rb atoms
formed in a magneto-optical trap (MOT). It is necessary to use cold
atoms to ensure that only the four intended atomic states are
excited, in contrast to a Doppler broadened room temperature gas. A
Kerr lens mode-locked Ti:Sapphire laser operating with an
approximately 55 nm bandwidth centered at 778 nm with $f_r \approx$
100 MHz is used to excite all four transitions. $f_r$ is phase
stabilized to a low phase-noise crystal oscillator, and steered via
a Cesium reference to maintain the absolute frequency of the comb
modes. The offset frequency $f_o$ is measured via a f-2f nonlinear
interferometer \cite{Cundiff1} and stabilized to a direct digital synthesizer. Regardless of the spectral phase of the
pulses, any comb mode has an absolute frequency given by,
$\nu_{\small N}=f_o+N\times f_r$, where $N$ is the mode order number (of
order $10^6$ for our laser). Using prior knowledge of the $^{87}$Rb
energy level structure for the 5S, 5P, and 5D hyperfine states, it
is possible to select a particular $f_r$ and $f_o$ to approximate a
diamond configuration with only four resonant levels (see Fig.
1(a)). We use two values of $f_r$, the first is $f_r$=100.59660605
MHz with $f_o$=+16.94 MHz. In this case, the four resonant states
are: 5S$_{1/2}$F=2, 5P$_{1/2}$F=2, 5P$_{3/2}$F=2, and 5D$_{3/2}$
F=1. With this selection of comb frequencies the transitions from
5S$_{1/2}$F=1, the other ground state, are at least 6 MHz detuned
from any intermediate and excited states. All other possible
transitions are further detuned. Note that the 5S to 5P and 5D linewidths are 6 MHz and 0.66 MHz respectively. Using a slightly shifted $f_r$ of
100.59660525 MHz and the same $f_o$, the comb is resonant with the
same intermediate and final states but from 5S$_{1/2}$ F=1. We
indirectly measure the 5D population by counting photons from the
5D-6P-5S cascade fluorescence at 420 nm with a photomultiplier tube.
The experimental cycle consists of three parts: first a MOT is
formed for 6.5 ms, then the atoms are held in optical molasses for 3
ms while the magnetic field turns off, finally the atoms are excited
for 0.5 ms and the photon counts are recorded versus time on a
multichannel scaler. We use a standard 2f-2f configuration pulse
shaper with a computer controlled spatial light modulator (SLM) to
set the relative phase of the comb modes with a per pixel
resolution of $\sim$150 GHz \cite{Weiner1}. The spatial and
temporal frequency chirp of the pulses at the MOT location are
reduced to maximize the fringe visibility.

\begin{figure}[b]
\resizebox{8.0cm}{!}{
\includegraphics[angle=0]{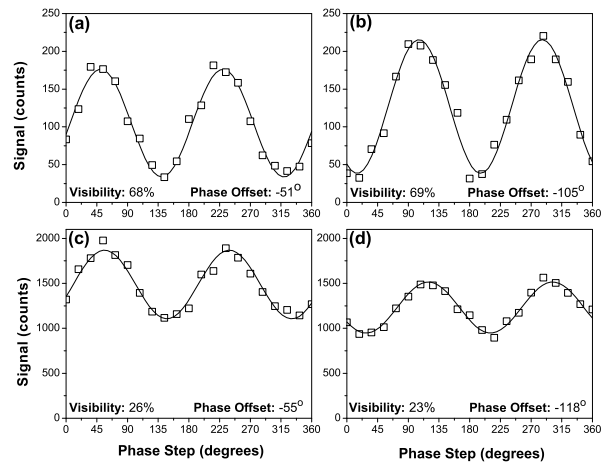}}
\caption{\label{Fig2}$\bf(a)$ Measured interference fringes with fits under
four different excitation conditions. All the results are obtained
by scanning the phase $\Phi$ of the phase mask shown in Fig. 1(b).
The top panels correspond to illuminating the atoms from only one
direction (traveling waves). The bottom panels are under
counter-propagating pulse excitation. The left panels (a) and (c)
use the $f_{r}$ for excitation from the 5S$_{1/2}$F=2 ground state;
the right panels (b) and (d) use the $f_r$ for excitation
from the 5S$_{1/2}$ F=1 ground state.}
\end{figure}

The first experiment provides a clear demonstration of the phase
sensitive nonlinear response of the four-level diamond
configuration. We use the spatial light modulator to apply the phase
mask indicated in Fig. 1(b), the effect of this mask is to change
the phase of all the comb mode pairs that are resonant from the
ground state to 5D$_{3/2}$F=1 and close to the 5P$_{3/2}$
intermediate states (denoted by the hatched region in Fig. 1(b)).
Specifically, the mask applies a variable phase step of $\Phi$ to
the spectral region from 772 nm to 784 nm. Due to the aforementioned
cancellation of the off-resonant amplitudes, it is sufficient to
consider only the mode pairs tuned nearest to an intermediate state
resonance for this first experiment.

Our measured results are shown in Fig. 2, with each fringe fit to a function of the form,
$\rho_{\small{5D}}=c_1+c_2 cos(\Phi+c_3)^2$, where $\Phi$ represents the
phase applied to the SLM, $c_3$ is a static phase offset, and the
fringe visibility is given by $\frac{c_2}{2c_1+c_2}$. Due to the
fact the phase mask covers both 780 nm and 776 nm, the two-photon
amplitude from the resonant path through 5P$_{3/2}$F=2 is phase
shifted by 2$\Phi$ and therefore has a period of $\pi$ radians. The
background counts due to ambient light at 420 nm and excitation of
the hot Rb atoms not trapped in the MOT are measured by repeating
the experiment without a MOT and subtracted from the reported data.
As mentioned previously, we choose to conduct this experiment with
two different values of $f_r$. The left panels in Fig. 2 are
measured with $f_r$ set for two-photon transitions from the
5S$_{1/2}$F=2 ground state. The right panels are with the second
$f_r$, resonant from the 5S$_{1/2}$F=1 ground state. For each $f_r$,
we also measure the interference fringe under excitation from a
single pulse propagation direction (Fig. 2 top panels) and under
counter-propagating pulses (Fig. 2 bottom panels).

\begin{table}[t]
\resizebox{8.0cm}{!}{
\includegraphics[angle=0]{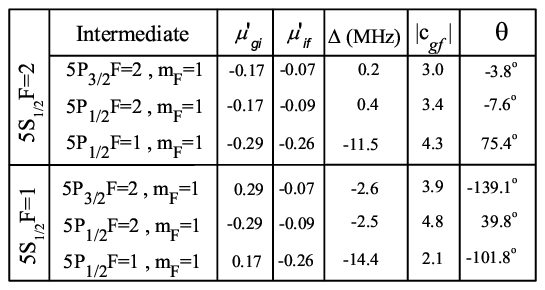}}
\caption{\label{Table} Left column is the intermediate state for each two-photon transition with a resonant or near-resonant comb mode. Across the top are: the reduced dipole moments, the detuning of the nearest mode from the intermediate state, the relative magnitude of the two-photon amplitude, and the phase of the amplitude. The top section is for transitions from 5S$_{1/2}$F=2 and the bottom from 5S$_{1/2}$F=1.}
\end{table}

Much like an optical interferometer, two important features of our
results are the fringe visibility and the phase offset. Clearly the
results presented in Fig. 2 show significant differences in both the
fringe visibility and phase offset as a function of excitation
scheme. We begin by discussing the offset of the fringe from
$\Phi$=0. Due to the large separation of wavelengths used in this
experiment and the significant dispersion of the pulse shaper and
other optics, any residual chirp of the pulse can cause an
overall phase shift common to all measured fringes. However, not all
fringes are shifted by the same amount; in particular, excitation
from the two different ground states yields results significantly
out of phase. To explain this relative phase shift we refer to Table 1, in which the key parameters to estimate the total
fringe visibility and phase shift are tabulated. The first and most
dominant effect is due to the sign of the dipole matrix elements;
for transitions from 5S$_{1/2}$F=2 all the dipole moments are
negative, however, for transitions from 5S$_{1/2}$F=1 there is sign
difference between different two-photon paths. Table 1 gives the
angular part of the dipole matrix elements, $\langle
Lm_FF||\hat{\bold r}||L'm_F'F'\rangle$, denoted as $\mu'_{gi}$ and
$\mu'_{if}$. Clearly the sign of the matrix elements affect the
interference in closed loop excitation, an important consideration for phase-resolved two-dimensional spectroscopy \cite{Jonas1}.

It is also necessary to consider the phase shift of a particular
two-photon amplitude due to the detuning from the relevant
intermediate state. For this we include in Table 1 the path through
the additional intermediate state 5P$_{1/2}$F=1, although the
nearest comb mode is detuned, the dipole moment is sufficiently
large to make its contribution significant. Due to the detuning of
this transition path, there is a large phase shift of the
corresponding amplitude. The effect of this additional path through
5P$_{1/2}$F=1 is to phase shift the total transition amplitude via
5P$_{1/2}$ relative to the 5P$_{3/2}$ amplitude, and thus the fringe
shift. This occurs for two-photon transitions from both ground
states. However, as can be seen from the dipole moments and
amplitudes in Table 1, the effect of the 5P$_{1/2}$F=1 state is less
for the 5S$_{1/2}$F=1 ground state case. The difference in fringe shift
between Fig. 2(a) and (b), corresponding to excitation from the two
ground states, is 54$^o$. This is in good agreement with the theoretically predicted value of
56$^o$.

The second feature, the fringe visibility, is a result of the
coherent interference between the 5P$_{1/2}$ and 5P$_{3/2}$ paths in
the diamond configuration and any additional incoherent signal which
raises the measured fringe minimum. Referring to the results in Fig.
2, the fringe visibility is strongly reduced under standing wave
excitation and exhibits little dependence on the choice of ground
state. In the case of traveling wave excitation, all the atoms in
the MOT are excited by the same relative magnitude of electric
fields. The visibility predicted using the amplitudes presented in
Table 1 is 82\% for excitation from the 5S$_{1/2}$F=2 ground state
and 92\% for the 5S$_{1/2}$F=1 ground state, assuming equal
population distribution among the $m_F$ sublevels. The best
experimental results obtained for the visibility are about 70\%,
shown in Fig. 2(a) and (b). Residual frequency and spatial chirps
have likely lowered the observed visibility from the ideal case. For
traveling wave excitation the interference effect is observed for
only the first 50 $\mu$s of excitation after which the atoms are
Doppler shifted completely off of resonance due to radiation pressure. It is for this reason
the data presented in Fig. 2 is only the first 10 $\mu$s of
excitation; using a larger time window significantly reduced the
fringe visibility.

One method to reduce the effect of radiation pressure on the atoms is to balance the average force by probing with counter-propagating pulses ~\cite{Stowe1}. The bottom panels of Fig. 2 present the
fringe measured using well overlapped
counter-propagating beams of the same intensity. Although we measure
a constant fringe visibility in this case for an extended time of
300 $\mu$s, the multi-mode standing wave generated by the
counter-propagating pulses reduces the visibility to $\sim$25\%.
This effect arises because the four main resonance frequencies have
different wavelengths and thus different standing-wave periods, so
the magnitudes of the four resonant electric fields vary spatially.
For example, in some regions of the atom cloud the 780 nm field
is maximum while the 794 nm field is minimum, and there is no
interference effect. In this case a spatial average over the atom
cloud, taking into account the different standing waves, must be
conducted. Using the amplitudes given in Table 1, the visibility
under multimode standing wave excitation is 36\% and 44\% from the
F=2 and F=1 ground states, respectively.

\begin{figure}[t]
\resizebox{8.50cm}{!}{
\includegraphics[angle=0]{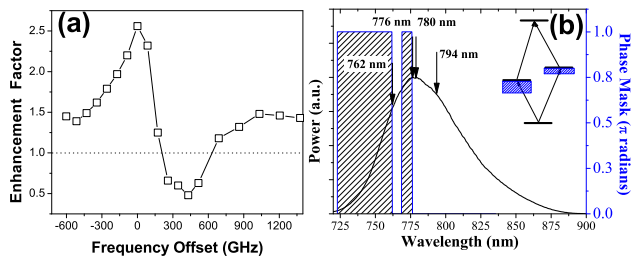}}
\caption{\label{Fig3} $\bf(a)$ Measured signal enhancement. The ratio of
signals with and without the phase mask in (b), versus position of
the SLM in the spectrum. The zero of the offset frequency is chosen
to be the position of the maximum signal increase. $\bf(b)$ The
applied phase mask is indicated by the hatched region. Exactly $\pi$
radians of phase is applied just below 762 nm and 776 nm to add an
extra phase shift to those mode pairs that join in the hatched
region.}
\end{figure}

The first experiment focuses entirely on those comb modes near
an intermediate resonance. This is due to the fact that the vast
majority of modes cancel out and thus make no net contribution to
the two-photon amplitude. In the second experiment, we use $f_r$ and $f_o$ for two-photon transitions from the 5S$_{1/2}$F=2 ground state and apply the phase mask presented in Fig. 3(b). This phase mask forces constructive interference between the amplitudes due to the off-resonant mode pairs, increasing the total signal. Recall for every two-photon resonant mode pair
detuned below an intermediate state, there is a pair detuned
approximately equally above the state. For a transform-limited pulse
train, these two pairs of modes give rise to excited state
amplitudes that are equal and opposite, and therefore cancel to
zero. By applying the phase mask in Fig. 3(b), those modes that are
detuned below either the 5P$_{3/2}$ or 5P$_{1/2}$ states (the
hatched area in Fig. 3(b)) obtain a 180$^o$ phase shift with respect
to those detuned above the intermediate state. This type of spectral
phase negates the inherent phase change due to the detuning around a
resonance (see Eq. 1), and causes constructive interference. Figure
3(a) shows we achieve a maximum increase of 2.56 over the normal
signal. The theoretical enhancement is 2.85, however, this estimate does not include the effects of diffraction at the phase steps in the SLM, which likely reduces the maximum. This data
is obtained by first coarsely tuning the position of the phase mask
at the per-pixel resolution. Then for finer control of the location
of the phase step applied to the comb spectrum, the entire SLM is
shifted using a micrometer through the spectrally dispersed optical
field.

In conclusion, we have demonstrated the precise control of a diamond configuration four-level atomic coherence over a 32 nm spectral width. The
first experiment focuses on the comb modes resonant with
intermediate states, and the second optimizes the two-photon
transition rate using high-resolution spectral phase shaping
to force constructive interference between the off-resonant modes.
The demonstrated ability in high-resolution control of the coherence
of a four-level system over a very broad-bandwidth may enable future
research in nonlinear optics of multi-level systems. For example,
one proposal for a cw-VUV laser utilizes lasing without inversion in
a four-level diamond configuration \cite{Fry1}.

We acknowledge funding support from NSF, NIST, ONR, and DARPA.

\end{document}